\documentclass[runningheads]{llncs}
\usepackage[T1]{fontenc}
\usepackage{graphicx}
\usepackage{hyperref}
\usepackage{booktabs}
\usepackage{tikz}
\usepackage{caption}
\usepackage{subcaption}
\usepackage{pgf-pie}
\usepackage{amsmath}

\usepackage{cleveref}
\definecolor{red}{RGB}{230,10,30}

\begin{document}
%
\title{Data Diet: Can Trimming PET/CT Datasets Enhance Lesion Segmentation?}
%
%
\author{Alexander Jaus\inst{1,2}\orcidID{0000-0002-0669-0300} \and
Simon Reiß\inst{1}\orcidID{0000-0003-1953-6211}  \and 
Jens Kleesiek \inst{3,4}\orcidID{0000-0001-8686-0682} \and
Rainer Stiefelhagen\inst{1}\orcidID{0000-0001-8046-4945}}
\authorrunning{A. Jaus et al.}
%
\institute{Karlsruhe Institute of Technology, Karlsruhe, Germany \and
HIDSS4Health - Helmholtz Information and Data Science School for Health, Karlsruhe/Heidelberg, Germany \and
Institute for AI in Medicine, University Hospital Essen, Essen, Germany \and 
Cancer Research Center Cologne Essen (CCCE), University Medicine Essen, Essen, Germany
}
\maketitle              
\begin{abstract}
In this work, we describe our approach to compete in the autoPET3 datacentric track. While conventional wisdom suggests that larger datasets lead to better model performance, recent studies indicate that excluding certain training samples can enhance model accuracy. We find that in the autoPETIII dataset, a model that is trained on the entire dataset exhibits undesirable characteristics by producing a large number of false positives particularly for PSMA-PETs. We counteract this by removing the easiest samples from the training dataset as measured by the model loss before retraining from scratch. Using the proposed approach we manage to drive down the false negative volume and improve upon the baseline model in both false negative volume and dice score on the preliminary test set. Code and pre-trained models are available at: \href{https://github.com/alexanderjaus/autopet3_datadiet}{github.com/alexanderjaus/autopet3\_datadiet}
\end{abstract}
\section{Introduction and Motivation}
The segmentation of cancer and other diseases in multimodal medical images, including CT, MRI, X-ray, and PET, is a long-standing challenge in the field of medical image analysis~\cite{heller2021state},~\cite{simpson2019large},~\cite{antonelli2022medical}. The autoPET challenge~\cite{gatidis2023autopet} aims to address the segmentation of lesions in a PET/CT dataset and is in its third iteration.
This third iteration brings two novelties: (1) the dataset contains FDG-PET and PSMA-PET allowing participants to test their models on two different tracers and (2) participants may compete in two tracks, the standard- and the datacentric track.
While wide-ranging modifications to the model architecture, used datasets, loss functions, and training pipeline are allowed in the standard track, the datacentric track keeps the model and optimization procedure fixed.
As such, in this track, only specific modifications to the dataset and training pipeline are permitted.
In particular, no external data may be used, alterations to augmentation strategies in the training pipeline are permitted while minor changes in the training setup, \textit{e.g.} altering the number of training epochs, are allowed.


Inspired by the spirit of this challenge, we embrace a data-centric approach and ask: \emph{Do we really need all annotated volumes in training?}
Generally, the mantra in deep learning is `more data is better', while this sentiment seems to be true for the most part~\cite{radford2021learning} it does not apply without restriction as recent research on excluding thoughtfully selected samples in training~\cite{gadre2024datacomp} shows. 
We deem approaches that weed out data which obfuscates training, to align closer to the principle of data-centrism as opposed to employing advanced data augmentation techniques~\cite{zhang2017mixup,ghiasi2021simple,yun2019cutmix} or enhancing performance through scaling training epochs and hardware for an increased batch size. 
To this effect, we do not change a single line of code from the baseline source-code~\footnote{\url{https://github.com/ClinicalDataScience/datacentric-challenge}} and solely control which samples are used in training.


\section{Method}
In the following section, we first describe the baseline model and then highlight its weaknesses.
Subsequently, we provide reasoning behind our data-centric filtering strategy with which we aim to address the weaknesses by sample exclusion.

\subsection{Baseline Segmentation Model} The autoPET authors present the DynUnet~\cite{isensee2019automated} as their baseline model of choice following the well-known setup of nnU-Net~\cite{isensee2021nnu}. The setup of the fixed model hyperparameters can be found in~\Cref{Tab:Fixed Model Params}.

\begin{table}[ht]
\centering
\renewcommand{\arraystretch}{1} 
\setlength{\tabcolsep}{5pt} 
\begin{tabular}{l|l|l|l|l}
\toprule
\textbf{Architecture} & \textbf{Patch Size} & \textbf{Spacing} & \textbf{Kernels} & \textbf{Strides} \\ \midrule
DynUnet~\cite{isensee2019automated} & (128, 160, 112) & (2.036, 2.036, 3.0) & $\{[3, 3, 3]\} \times 6$ & $\{[1, 1, 1]\}\times 1$, \\
                                     &                 &                    &                      & $\{[2, 2, 2]\}\times 4$, \\
                                     &                 &                    &                      & $\{[1, 2, 1]\}\times 1$ \\ 
\bottomrule
\end{tabular}
\caption{Model parameters of the fixed baseline DynUnet model for the data-centric autoPET challenge track.}
\label{Tab:Fixed Model Params}
\end{table}

\noindent The basic training pipeline used by the authors of this baseline includes training the model on the full autoPET dataset.
More specifically, normalization is used to reorient the images into a $LAS$ coordinate framework, resample to a common spacing, and bring the CT and PET values to a fixed range of $[0,1]$.
Further, the basic training makes use of data augmentation techniques such as introducing gaussian noise, gaussian smoothing, simulating low-resolution images, randomly adjusting image contrast, and flipping along all three axes.
For reproducibility, a deterministic setup is used through fixing random seeds.


\subsection{Analyzing the Baseline}
\begin{figure}
    \centering

    \begin{tabular}{cc} 
        \includegraphics[width=0.5\textwidth, trim=50 0 0 0,clip]{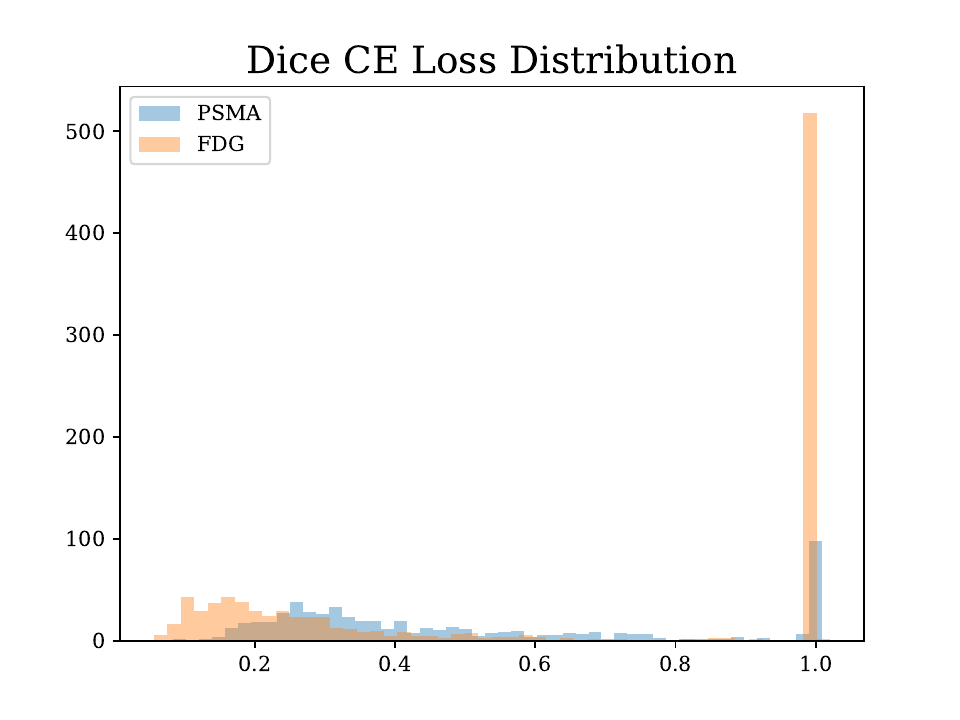} & 
        \includegraphics[width=0.5\textwidth, trim=50 0 0 0,clip]{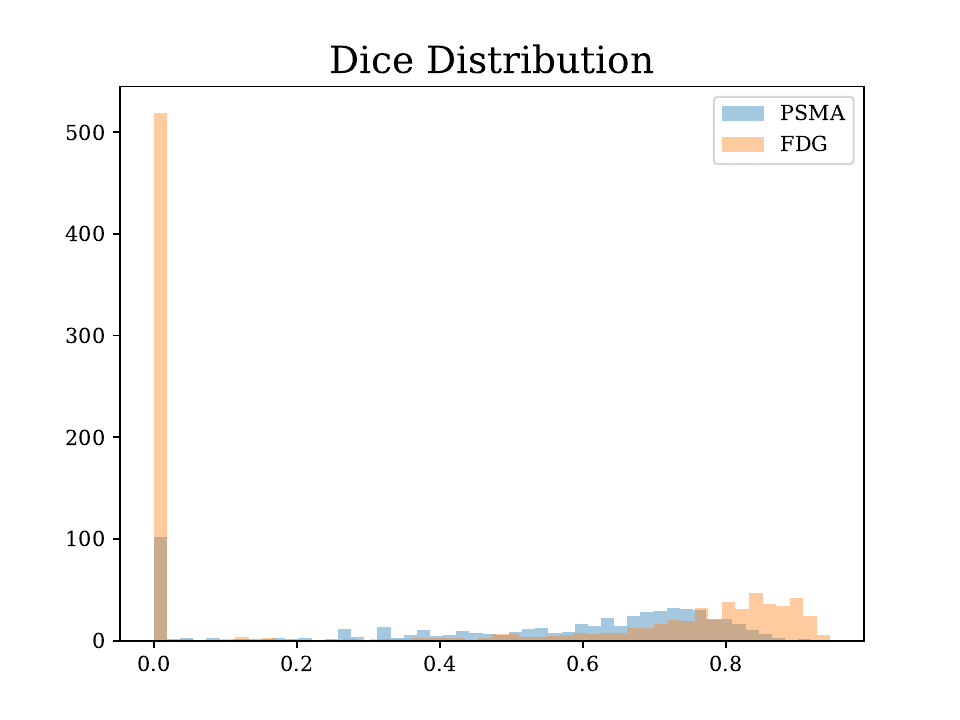} \\
        \includegraphics[width=0.5\textwidth, trim=50 0 0 0,clip]{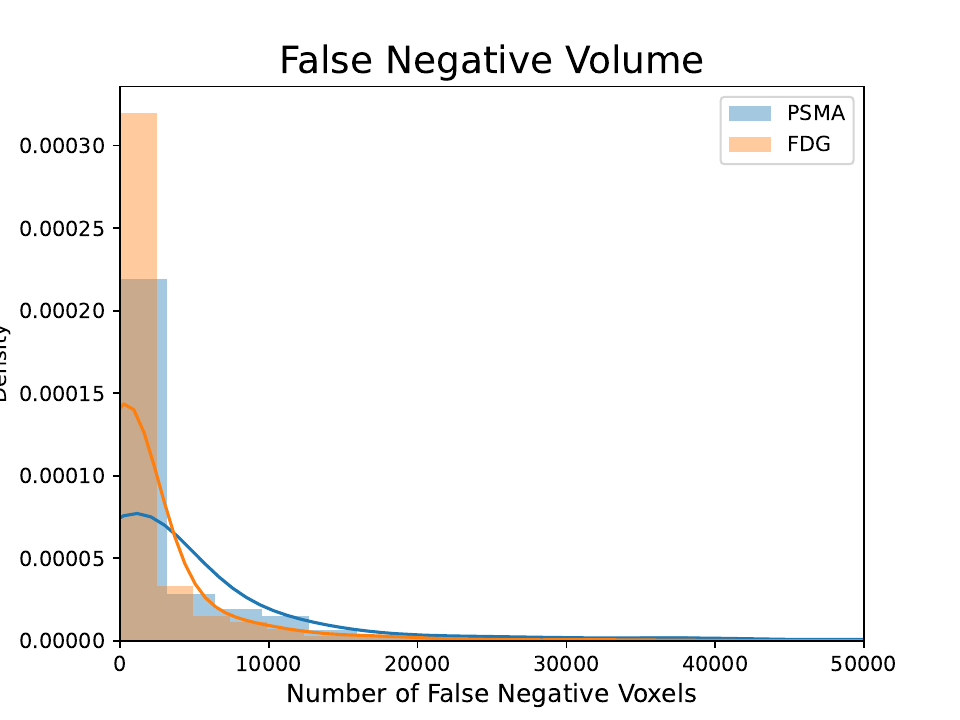} & 
        \includegraphics[width=0.5\textwidth, trim=50 0 0 0,clip]{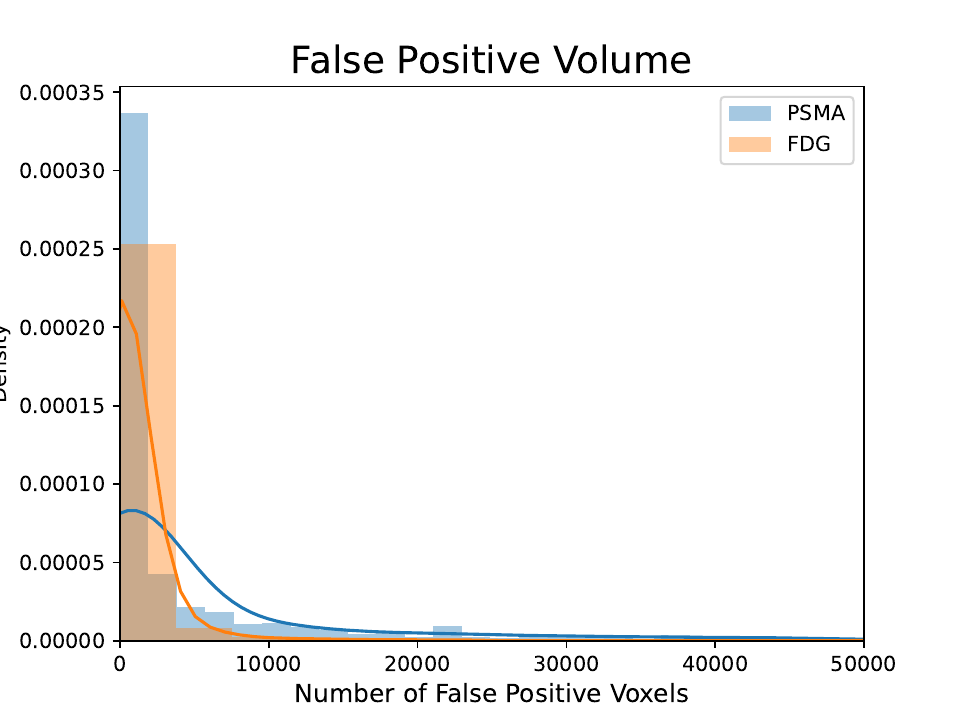} \\
    \end{tabular}

    \caption{Analysis of the baseline model on the entire autoPET training dataset}
    \label{fig: Baseline Analysis}
\end{figure}

As the authors provide the weights of the baseline model, we start by exploring its properties.
The model was trained on the entire training dataset using all folds, to characterize it, we evaluate it's performance on this same data.
Particularly, we measure the magnitude of the loss function, the Dice score, the false positive volumes, and the false negative volumes for each sample. These measurements in the form of histograms and regressed graphs, split into PSMA- and FDG-PETs are shown in~\Cref{fig: Baseline Analysis}.
Following insights can be drawn from these measurements:

(1) Generally, FDG-PETs are learned better than PSMA-PETs. This can be seen in both, a right-shifted distribution of the FDG-PET Dice scores in the upper right plot, and a left-shifted distribution of loss values in the upper left plot, which indicates better alignment of the learned and target distribution.

(2) The model tends to under-segment metastasis for both FDG- and PSMA-PETs. We find that, the PSMA  and FDG distribution of false negative volumes in the lower left plot roughly share the same shape, indicating that the baseline model tends to under-segment metastasis equally for FDG- and PSMA-PETs.

(3) PSMA-PETs exhibit a higher number of false positive voxels.
When comparing the distribution of false positive volumes, we find a slight difference in the distributions between PSMA and FDG tracers, where the false positive distribution of PSMA-PETs tends to be skewed more to the right compared to  FDG-PETs indicating that PSMA-PETs exhibit more false positive voxels. 

Putting these insights together, we hypothesize that the observed performance of the baseline model is to a large extent limited by its predictions in PSMA-PETs, leading to high false positive values.


\subsection{Analyzing the Dataset}
\begin{figure}
    \centering
    \begin{subfigure}[b]{0.3\textwidth}
        \centering
        \begin{tikzpicture}
            \pie[scale=0.4, color={teal, lightgray}]{62.9/FDG, 37.1/PSMA}
        \end{tikzpicture}
        \caption{Tracers}
    \end{subfigure}
    \begin{subfigure}[b]{0.3\textwidth}
        \centering
        \begin{tikzpicture}
            \pie[scale=0.4, color={teal, lightgray}]{49.4/Sick, 51.6/Healthy}
        \end{tikzpicture}
        \caption{FDG: Sick Rate}
    \end{subfigure}
    \begin{subfigure}[b]{0.3\textwidth}
        \centering
        \begin{tikzpicture}
            \pie[scale=0.4, color={teal, lightgray}]{90/Sick, 10/Healthy}
        \end{tikzpicture}
        \caption{PSMA: Sick Rate}
    \end{subfigure}
    \caption{Comparison of the number of samples for each tracer within the autoPet training dataset and the morbidity ratio for each tracer.}
    \label{fig:PieDiagrams}
\end{figure}
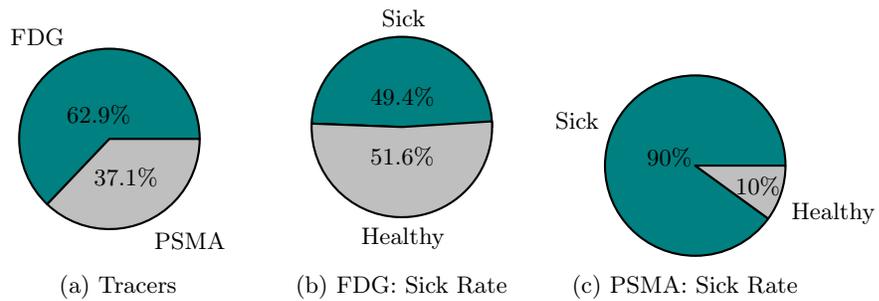

Upon investigation of the dataset, we find that there is some form of imbalance between the FDG-PET and the PSMA-PET samples. The total number of images in the training dataset is $1,611$ of which $1,014$ are FDG-PET and $597$ are PSMA-PET images.
Further, the ratio between sick and healthy patients for each tracer are quite balanced for FDG-PETs but severely imbalanced for PSMA PETs, as seen in~\Cref{fig:PieDiagrams}.
This clue supports our stated hypothesis that the baseline tends to overconfidently predict metastasis in PSMA-PET images, since it is trained with only few healthy PSMA-PETs.
Thus, the model opts to predict critical regions as cancer, rahter than not.

\subsection{Proposed Method}

According to our previous analysis, we find, that an ideal, data-centric solution should focus on reducing the predicted false positive volumes of the model for PSMA-PETs.
With our intent of selecting samples to exclude from the training set, we reduce the search space for such samples to the PSMA-PETs and keep all FDG-PET images as FDG tracer images scored already quite high in our measurements.
Regarding the PSMA-PETs we exclude images for which the baseline model produces a \emph{very low loss}, which are potential candidates that tilt the model towards overconfidence, as the model overfits on them.
To this end, we use the measured loss from the previous step and order the PSMA-images by their loss magnitude in ascending order, then we prune the $n^{\text{th}}$ percentile from the training dataset before retraining from scratch, where $n$ is a hyperparameter that we ablate. When analyzing the PSMA images that produce very low loss values ($n\leq 5$), we find not a single image of them contains a healthy patient.
This approach that we call data diet poses two advantages: (1) It improves the sick-to-healthy ratio of the training dataset and (2) hard examples -- where learning happens -- remain in the training dataset which should lead to a better-calibrated model and thus reduce the false positives.

\section{Results}
As defined by the challenge, we copy the exact setup of the baseline model and do not make any changes, neither to the model nor to the training pipeline. In order to compare our setup directly to the baseline, we use the preliminary test set of the challenge for evaluation. We vary the percentile of easiest PSMA examples $n$ and measure the model performance. We summarize our results in~\Cref{Tab:Resuls}.

\begin{table}[ht]
\centering
\renewcommand{\arraystretch}{1.3} 
\setlength{\tabcolsep}{5pt} 
\begin{tabular}{c|c|c|c}
\toprule
\textbf{Excluded Percentile} & \textbf{Dice} & \textbf{FN Volume} & \textbf{FP Volume} \\ \midrule
Baseline & 0.7990 & 23.8380 & \textbf{2.0503} \\
$1^{\text{st}}$ Percentile                 & 0.7879 & 24.3688 & 2.7718 \\
$3^{\text{rd}}$ Percentile                  & \textbf{0.8129} & 18.8968 & 2.1739 \\
$4^{\text{th}}$ Percentile                                   & 0.7569 & 20.8366 & 2.3273 \\
$5^{\text{th}}$ Percentile                                   & 0.7980 & \textbf{17.3220} & 8.2732 \\
\bottomrule
\end{tabular}
\caption{Results on the preliminary testset, we show the best results in \textbf{bold}.}
\label{Tab:Resuls}
\end{table}

\begin{figure}
    \centering
    \includegraphics[width=0.66\linewidth, trim=20 5 40 38,clip]{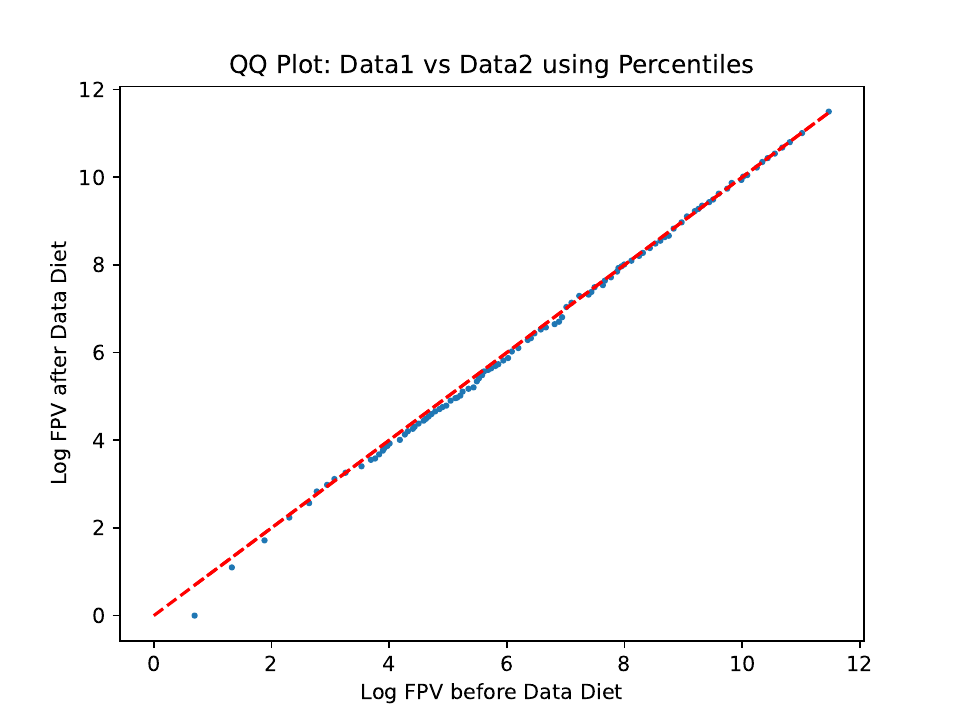}
    \caption{QQ-Plot of the pre vs. post data diet False Positive volumes distribution. The dashed red line is the reference indicating an equal distribution.}
    \label{fig:qq-plots}
\end{figure}

Taking a look at the Dice score, which is driven by both, false positives as well as false negatives, we find the best results when excluding $3\%$ of the easiest PSMA images, we however find that due to the limited size of the preliminary testset (only 5 images) we can neither confirm nor contradict our proposed strategy. We give the full list of excluded images for the $5\%$ and the $3\%$ scenario in~\Cref{sec:ex}. To further evaluate whether the proposed approach reduced the number of false positive predictions in the PSMA-Tracer, we generated a QQ-plot of the log-percentiles from the PSMA False-Positive Distribution, comparing the training dataset before and after a 3\% data diet. The results are presented in~\cref{fig:qq-plots}. The plot reveals that most quantiles of the distribution shifted below the dashed reference line, which indicates that the pre- and post-diet distributions are equal. This shift demonstrates a reduction in false positive values, supporting the effectiveness of the data diet approach. Interestingly, the data diet did not alter the False-Negative Volume distribution, as shown in a similar log QQ-plot in~\Cref{qq-plot-fnv}.

\section{Discussion and Limitation}
While the overall hypothesis seems to match the observed results it has to be noted that due to the rather small size of the preliminary test set consisting only of a handful of patients, further experiments would be required to confirm the empirical evidence. 
Further, we rely on the assumption that modifying the PSMA-data does not influence the performance of the FDG-PET images which would require additional experiments to be validated or falsified.

\section{Acknowledgements}
The present contribution is supported by the Helmholtz Association under the joint research school “HIDSS4Health – Helmholtz Information and Data Science School for Health and was supported by funding from the pilot program Core-Informatics of the Helmholtz Association (HGF). This work was performed on the HoreKa supercomputer funded by the Ministry of Science, Research and the Arts Baden-Württemberg and by the Federal Ministry of Education and Research.

%
%
\bibliographystyle{splncs04}
\bibliography{samplepaper}

\appendix
\newpage
\section{Excluded Patients in $5^{\text{th}}$ percentile}
\label{sec:ex}
\begin{verbatim}

psma_324f91cd0ec8a80e_2017-03-04
psma_ec45934c2fa23c76_2019-08-05
psma_ec55aac3ff6bec15_2017-04-24
psma_709da54824f16bc1_2018-02-19
psma_a2ff166a8aad6c9f_2018-03-01
psma_5095b111f4d06a37_2017-02-03
psma_25c60014bb1469d3_2022-05-14
psma_698f9e802d84253c_2014-08-03
psma_a672c1e6a35c21b8_2019-07-20
psma_9a8d486e8ddd0427_2021-03-20
psma_c408e60627628585_2021-11-19
psma_4c9d9614d81f3005_2019-10-04
psma_5813d6984082108d_2016-10-23
psma_a17bb16aa6d2f6fa_2020-10-31
psma_8f09d57384f77ce6_2017-07-31
psma_4c9d9614d81f3005_2019-04-15
psma_7e25beff54698eb8_2021-02-06
psma_56e540714db4775a_2020-03-27
psma_18f866bfa3d793d4_2016-08-01
psma_ba5034e13e4c31f5_2019-07-05
psma_6d5c002d629eb131_2022-01-22
psma_0ec0e718244b2a79_2018-09-28
psma_28b47ab366f7ec9d_2016-04-04
psma_ffcaa75377465b37_2018-03-04
psma_80a782c4b998fefd_2020-07-31
psma_57a5841a095d55d1_2018-06-30
psma_95a4d87af2bc56f7_2019-03-11
psma_7fa68d5b0861c0b4_2014-10-30
psma_ffcaa75377465b37_2017-07-20
psma_ffcaa75377465b37_2016-11-14
    
\end{verbatim}

\section{Excluded Patients in $3^{\text{rd}}$ percentile}
\begin{verbatim}
psma_5813d6984082108d_2016-10-23
psma_a17bb16aa6d2f6fa_2020-10-31
psma_8f09d57384f77ce6_2017-07-31
psma_4c9d9614d81f3005_2019-04-15
psma_7e25beff54698eb8_2021-02-06
psma_56e540714db4775a_2020-03-27
psma_18f866bfa3d793d4_2016-08-01
psma_ba5034e13e4c31f5_2019-07-05
psma_6d5c002d629eb131_2022-01-22
psma_0ec0e718244b2a79_2018-09-28
psma_28b47ab366f7ec9d_2016-04-04
psma_ffcaa75377465b37_2018-03-04
psma_80a782c4b998fefd_2020-07-31
psma_57a5841a095d55d1_2018-06-30
psma_95a4d87af2bc56f7_2019-03-11
psma_7fa68d5b0861c0b4_2014-10-30
psma_ffcaa75377465b37_2017-07-20
psma_ffcaa75377465b37_2016-11-14
\end{verbatim}
\section{Comparison of False Negative Distribution Pre- vs. Post Data Diet}
\label{qq-plot-fnv}
\begin{figure}
    \centering
    \includegraphics[width=0.66\linewidth, trim=20 5 40 38,clip]{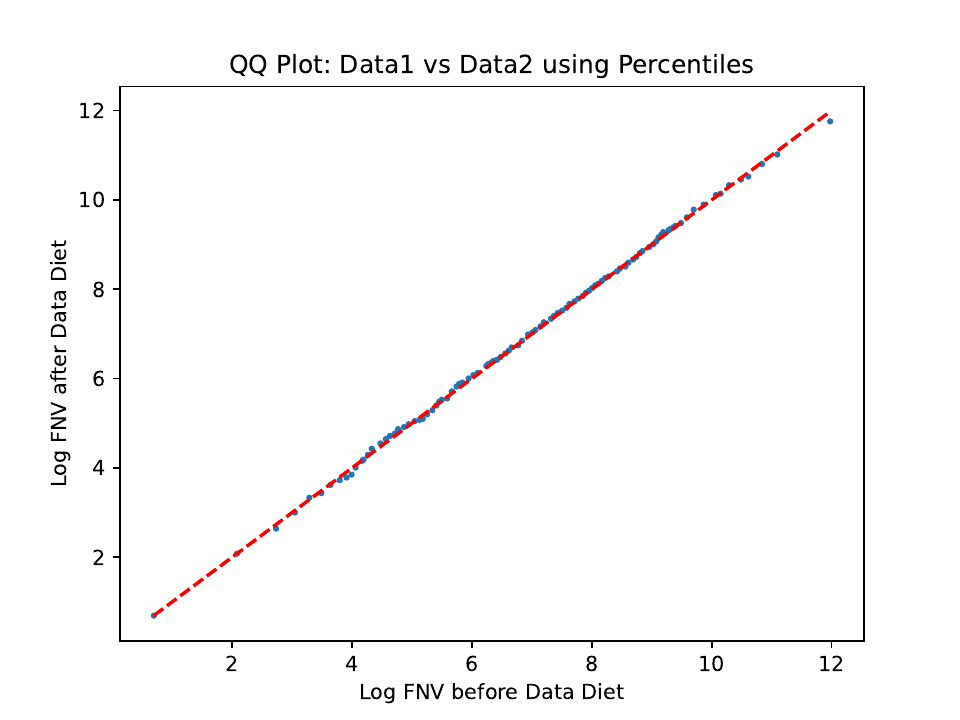}
    \caption{QQ-Plot of the pre-vs. post diet False-Negative Volume distribution. The dashed red line is the reference indicating an equal distribution.}
\end{figure}
\end{document}